\documentclass[prd, showpacs,preprintnumbers,nofootinbib]{revtex4}
\usepackage[dvips]{graphicx}
\usepackage{enumerate}
\usepackage{amsmath,amssymb}
\usepackage{mathrsfs}
\usepackage[dvips]{graphicx}
\usepackage{bm}

\begin{document}
\preprint{KEK Preprint 2012-36/CHIBA-EP-199, 2012}

\title{Non-Abelian dual superconductivity in SU(3) Yang-Mills theory: 
\\
dual Meissner effect and type of the vacuum
}

\author{Akihiro Shibata$^{1}$}
\email{akihiro.shibata@kek.jp}

\author{Kei-Ichi Kondo$^{2}$}
\email{kondok@faculty.chiba-u.jp}

\author{Seikou Kato$^{3}$}
\email{skato@fukui-nct.ac.jp}

\author{Toru Shinohara$^{2}$}
\email{sinohara@graduate.chiba-u.jp}

\affiliation{
$^1$Computing Research Center,  High Energy Accelerator Research Organization (KEK),  
and Graduate Univ. for Advanced Studies (Sokendai), Tsukuba  305-0801, Japan
\\
$^2$Department of Physics,  
Graduate School of Science, 
Chiba University, Chiba 263-8522, Japan
\\
$^3$Fukui National College of Technology, Sabae 916-8507, Japan
}

\begin{abstract}
We have proposed the non-Abelian dual superconductivity picture for quark confinement in the $SU(3)$ Yang-Mills (YM) theory, and have given numerical evidences for the restricted-field dominance and the non-Abelian magnetic monopole dominance in the string tension by applying a new formulation of the YM theory on a lattice.
To establish the non-Abelian dual superconductivity picture for quark confinement, we have observed the  non-Abelian dual Meissner effect in the $SU(3)$ Yang-Mills theory by measuring the chromoelectric flux  created by the quark-antiquark source, and the non-Abelian magnetic monopole currents induced around the flux.
We conclude that  the dual superconductivity of the $SU(3)$ Yang-Mills theory is strictly the type I and that this type of dual superconductivity is reproduced by the restricted field and the non-Abelian magnetic monopole part, in sharp contrast to the $SU(2)$ case: the border of type I and type II. 

\end{abstract}


\maketitle

\section{Introduction}

Quark confinement follows from the area law of the Wilson loop average. The dual superconductivity is the promising mechanism for quark confinement \cite{dualSC}. 
In many preceding works, the Abelian projection \cite{tHooft81} was used to perform numerical analyses, which exhibited the remarkable results such as Abelian dominance \cite{Suzuki90}, magnetic monopole dominance \cite{stack94,shiba}, and center vortex dominance \cite{greensite} in the string tension. 
However, these results are obtained only in special gauges: the maximal Abelian (MA) gauge and the Laplacian Abelian gauge within the Abelian projection, which  breaks the gauge symmetry as well as color symmetry (global symmetry).

In order to overcome the shortcomings of the Abelian projection,  we have presented a new lattice formulation of $SU(N)$ Yang-Mills (YM) theory in the previous papers \cite{SCGTKKS08L,exactdecomp} (as a lattice version of the continuum formulations \cite{CFNS-C,KSM05} for $SU(2)$ and \cite{SCGTKKS08} for $SU(N)$), which  gives a decomposition of the gauge link variable suited for extracting the dominant modes for quark confinement in the gauge independent way. 
In the case of $SU(2)$, the decomposition of the gauge link variable was  given on  a lattice \cite{KKMSSI06,ref:NLCVsu2,ref:NLCVsu2-2,kato:lattice2009} as a lattice version of the Cho-Duan-Ge-Faddeev-Niemi   decomposition \cite{CFNS-C}.
For the gauge  group $G=SU(N)$ ($N \ge 3$), it was found that the extension of the decomposition from  $SU(2)$  to $SU(N)$ ($N \ge 3$) is not unique and that  there are a number of possible ways of decompositions discriminated by the stability subgroup $\tilde{H}$ of $G,$ while there is the unique option of $\tilde{H}=U(1)$ in the $SU(2)$ case \cite{kondo:taira:2000}.

 For the case of $G=SU(3)$, in particular, there are two possibilities which we call the maximal option and the minimal option. 
The maximal option is obtained for the stability group $\tilde{H}=U(1)\times U(1)$, which enables us to give a gauge invariant version of the MA gauge as the Abelian projection \cite{lattce2007,Suganuma,Cho80c,FN99a}. 
The minimal one is obtained for the stability group  $\tilde{H}=U(2)\cong SU(2)\times U(1)$, which is suited  for representing the Wilson loop in the fundamental representation as derived from the non-Abelian Stokes theorem \cite{KondoNAST,KondoShibata}.
In the static potential for a pair of  quark and antiquark in the fundamental representation, we have demonstrated in
\cite{lattice2008,lattice2009,lattice2010} and \cite{abeliandomSU(3)}:
(i) the  restricted-field dominance or ``Abelian''  dominance  (which is a gauge-independent (invariant) extension of the conventionally called Abelian  dominance):  the string tension $\sigma_{\rm V}$ obtained from the decomposed $V$-field (i.e., restricted  field) reproduced the string tension $\sigma_{\rm full}$ of the original YM field,
$\sigma_{\rm V}/\sigma_{\rm full}=93\pm16\%$,
(ii)  the gauge-independent non-Abelian magnetic monopole dominance:  the string tension $\sigma_{\rm V}$ extracted from the restricted field was reproduced by only the (non-Abelian) magnetic monopole part $\sigma_{\rm mon}$,
$\sigma_{\rm mon}/\sigma_{\rm V}=94\pm9\%$.

The purpose of this paper is to give further evidences for establishing the non-Abelian dual superconductivity picture for quark confinement in $SU(3)$ Yang-Mills theory   
claimed in \cite{abeliandomSU(3)} by applying our new formulation to the $SU(3)$ YM theory on a lattice. 
First, we study the  dual Meissner effect by measuring the distribution of chromo-flux created by a pair of static quark and antiquark.
We compare the chromo-flux of the original Yang-Mills field with that of the restricted field and examine if the restricted field corresponding to the stability group $\tilde{H}=U(2)$ reproduces the dual Meissner effect, namely, the dominant part of the chromoelectric field strength of $SU(3)$ Yang-Mills theory.
Second, we measure the possible magnetic monopole current induced around the flux connecting a pair of static quark and antiquark. 
Third, we focus on the type of dual superconductivity, i.e., type I or type II. 
In the $SU(2)$ case, the extracted field corresponding to the stability group $\tilde{H}=U(1)$ reproduces the dual Meissner effect, which gives a gauge invariant version of MA gauge in the Abelian projection, as will be given in \cite{DualSC:KKSS2012}. 
In this paper, we find that the dual superconductivity of the $SU(3)$ Yang-Mills theory is indeed the type I, in sharp contrast to the $SU(2)$ case: the border of type I and type II \cite{DualSC:KKSS2012}.

The chromo-flux of the Yang-Mills field has been already measured by using a gauge-invariant Wilson line/loop operator \cite{DiGiacomo:1990hc} by the preceding works:%
\cite{Dzhunushaliev:2004nm,Koma:2003hv,Koma:2003gq,Suzuki:2004dw,Chernodub:2005gz,Suzuki:2009xy,Matsubara:1993nq,Cea:2012qw} for $SU(2)$ case 
and \cite{Matsubara:1993nq,Cea:2012qw,Cardoso:2011cs,Cardaci2011,Bicudo:2010gv,Cardoso:2009kz,Cardoso:2009qt,Cardoso:2010cf,Cardoso:2010kw,Bakry:2011ew,Kamizawa:1992np,Cardoso:2012rb,Cardoso:2012aj,Cardoso:2012uk,Cardoso:2012rm,Bicudo:2011hq,Bicudo:2011hk,Cardoso:2011fq,Suganuma:2004hu,Ichie:2002mi,Mori:2003vs,Bornyakov:2003vx} for the $SU(3)$ case.
However, most of the $SU(2)$ case were done using the Abelian projection and there were no direct measurement of the dual Meissner effect in the gauge independent (invariant) way, except for quite recent studies  \cite{Cardaci2011,Cardoso:2011fq,Cea:2012qw}.

\section{Lattice formulation}

We give a quick review of a new formulation of the lattice $SU(N)$ YM theory \cite{SCGTKKS08L,exactdecomp}.
In particular, the minimal option of the formulation enables us to extract the dominant mode of quark confinement, if we focus our attention on quarks belonging to the fundamental representation of the $SU(N)$ gauge group \cite{abeliandomSU(3),lattice2010}. 
For the original $SU(3)$  gauge link variable $U_{x,\mu} \in SU(3)$,
we wish to decompose it into new variables $V_{x,\mu}$ and $X_{x,\mu}$ which have values in the $SU(3)$ group, i.e., $X_{x,\mu}  \in SU(3)$, $V_{x,\mu} \in SU(3)$: 
\begin{equation}
U_{x,\mu}=X_{x,\mu}V_{x,\mu} \in SU(3) ,
\end{equation} 
so that $V_{x.\mu}$ could be the dominant mode for quark confinement, while $X_{x,\mu}$ is the remainder. 
In this decomposition, we require that  $V_{x,\mu}$ is transformed in the same way as the original gauge link variable $U_{x,\mu}$ and $X_{x,\mu}$ as a site variable by the full $SU(3)$ gauge transformation $\Omega_{x}$:
\begin{subequations}
\label{eq:gaugeTransf}%
\begin{align}
U_{x,\mu}  &  \longrightarrow U_{x,\nu}^{\prime}=\Omega_{x}U_{x,\mu}%
\Omega_{x+\mu}^{\dag},\\
V_{x,\mu}  &  \longrightarrow V_{x,\nu}^{\prime}=\Omega_{x}V_{x,\mu}%
\Omega_{x+\mu}^{\dag},\text{ \ }X_{x,\mu}\longrightarrow X_{x,\nu}^{\prime
}=\Omega_{x}X_{x,\mu}\Omega_{x}^{\dag}.
\end{align}
\end{subequations}

First, we introduce the key variable $\mathbf{h}_{x}$ called the color field. 
In the minimal option, the color field is defined by
\begin{equation}
\mathbf{h}_{x}=\xi(\lambda^{8}/2)\xi^{\dag}   \in Lie[SU(3)/U(2)] ,
\end{equation}
with $\lambda^{8}$
being the Gell-Mann matrix and $\xi$ the $SU(3)$ group element.
Once the color field is introduced, the above decomposition is obtained by solving the defining equation:
\begin{align}
&  D_{\mu}^{\epsilon}[V]\mathbf{h}_{x}:=\frac{1}{\epsilon}\left[  V_{x,\mu
}\mathbf{h}_{x+\mu}-\mathbf{h}_{x}V_{x,\mu}\right]  =0 .
\label{eq:def1}
\end{align}

This defining equation  can be solved exactly
\cite{exactdecomp}, and the solution is given by
\begin{subequations}
\label{eq:decomp}%
\begin{align}
X_{x,\mu}  &  =\widehat{L}_{x,\mu}^{\dag}\det(\widehat{L}_{x,\mu})^{1/3}%
g_{x}^{-1},\text{ \ \ \ }V_{x,\mu}=X_{x,\mu}^{\dag}U_{x,\mu}=g_{x}\widehat
{L}_{x,\mu}U_{x,\mu},\\
\widehat{L}_{x,\mu}  &  =\left(  L_{x,\mu}L_{x,\mu}^{\dag}\right)
^{-1/2}L_{x,\mu},\\
L_{x,\mu}  &  =\frac{5}{3}\mathbf{1}+ \sqrt{\frac{4}{3}%
}(\mathbf{h}_{x}+U_{x,\mu}\mathbf{h}_{x+\mu}U_{x,\mu}^{\dag})+8\mathbf{h}%
_{x}U_{x,\mu}\mathbf{h}_{x+\mu}U_{x,\mu}^{\dag}\text{ .}%
\end{align}
\label{defeq}
\end{subequations}
Here the variable
$g_{x}$ is the $U(2)$ part undetermined from Eq.(\ref{eq:def1}) alone.
In what follows, therefore, we put 
\begin{align}
 g_{x} =1 ,
\label{eq:def2}%
\end{align}
so that the above defining equations (\ref{eq:def1}) and (\ref{eq:def2}) correspond  respectively to the continuum version \cite{SCGTKKS08}:
\begin{subequations}
\begin{align}
&D_{\mu}[\mathbf{V}]\mathbf{h}(x)=0, 
\\
& \mathbf{X}_{\mu}(x)-\frac43 [\mathbf{h}(x),[\mathbf{h}(x),\mathbf{X}_{\mu}(x)] = 0
 .
\end{align}
\end{subequations}
For the physical meaning of the defining equation, see \cite{KondoShibata}. 
In the naive continuum limit, indeed, (\ref{defeq}) reproduces  the  decomposition in the continuum theory \cite{SCGTKKS08}:
\begin{subequations}
\begin{align}
\mathbf{A}_{\mathbf{\mu}}(x) &=\mathbf{V} 
_{\mu}(x)+\mathbf{X}_{\mu}(x) ,
\nonumber\\
\mathbf{V}_{\mu}(x)  &  =\mathbf{A}_{\mathbf{\mu}}(x)-\frac{4}{3}\left[
\mathbf{h}(x),\left[  \mathbf{h}(x),\mathbf{A}_{\mathbf{\mu}}(x)\right]
\right]  -ig^{-1}\frac{4}{3}\left[  \partial_{\mu}\mathbf{h}%
(x),\mathbf{h}(x)\right]  ,\\
\mathbf{X}_{\mu}(x)  &  =\frac{4}{3}\left[  \mathbf{h}(x),\left[
\mathbf{h}(x),\mathbf{A}_{\mathbf{\mu}}(x)\right]  \right]  +ig^{-1}%
\frac{4}{3}\left[  \partial_{\mu}\mathbf{h}(x),\mathbf{h}(x)\right]  .
\end{align}
\end{subequations}

Thus the decomposition is uniquely determined  from    Eqs.(\ref{eq:decomp}), if the color field $\mathbf{h}_{x}$ is specified. 
In order to determine the configuration $\{\mathbf{h}_{x}\}$ of color fields, we use the reduction condition \cite{SCGTKKS08L,exactdecomp} which guarantees that  the new theory written in terms of  new variables ($X_{x,\mu}$,$V_{x,\mu}$) is equipollent to the original YM theory. 
Here, we use the reduction condition:
for a given configuration of the original link variables $U_{x,\mu}$, color fields $\left\{  \mathbf{h}_{x}\right\}  $ are obtained by minimizing the functional: 
\begin{equation}
F_{\text{red}}[\{ \mathbf{h}_{x} \}]=\sum_{x,\mu}\mathrm{tr}\left\{  (D_{\mu
}^{\epsilon}[U_{x,\mu}]\mathbf{h}_{x})^{\dag}(D_{\mu}^{\epsilon}[U_{x,\mu
}]\mathbf{h}_{x})\right\} .
 \label{eq:reduction} 
\end{equation}

From the non-Abelian Stokes theorem \cite{KondoNAST,KondoShibata} and the Hodge decomposition of the field strength $\mathcal{F}_{\mu\nu}[\mathbf{V}],$ we can define the magnetic part (magnetic monopole current) in the gauge independent way. 
The lattice version of the gauge-invariant magnetic-monopole current $k$  is given by using the $V$-field as
\begin{subequations}
\begin{align}
&  V_{x,\mu}V_{x+\mu,\nu}V_{x+\nu,\mu}^{\dag}V_{x,\nu}^{\dag}
=\exp\left(-ig \epsilon^2 \mathcal{F}_{\mu\nu}[\mathbf{V}](x) \right)  ,
\\
& \epsilon^2 \Theta_{\mu\nu}(x) :=-\arg\text{ \textrm{Tr}}\left[  \left(  \frac{1}%
{3}\mathbf{1}-\frac{2}{\sqrt{3}}\mathbf{h}_{x}\right)  V_{x,\mu}V_{x+\mu,\nu
}V_{x+\nu,\mu}^{\dag}V_{x,\nu}^{\dag}\right]  ,\\
&  k_{\mu}(x)=2\pi n_{\mu}(x) 
= \partial_{\nu} {}^*\Theta_{\mu\nu}(x) 
:=\frac{1}{2}\epsilon_{\mu\nu\alpha\beta}\partial_{\nu}\Theta_{\alpha\beta}(x) .
\end{align}
\end{subequations}
Note that the the current $k_{\mu}$ is the non-Abelian magnetic monopole
current, since $V_{x,\mu}$ is the decomposed restricted  field.

\section{Method and results}

We generate configurations of the YM gauge link variable $\{ U_{x,\mu} \}$ using the standard Wilson action on a $24^{4}$ lattice at $\beta=6.2$.
The gauge link decomposition is obtained according to the framework given in the previous section: the color field configuration  $\{ h_{x} \}$ is obtained by solving the reduction condition of minimizing the functional eq.(\ref{eq:reduction}) for each gauge configuration  $\{ U_{x,\mu} \}$, and then the decomposed variables  $\{ V_{x,\mu} \}$,  $\{ X_{x,\mu} \}$ are obtained by using the formula eq.(\ref{eq:decomp}). 
In the measurement of the Wilson loop average, we apply the APE smearing technique to reduce noises \cite{APEsmear}.

\subsection{Dual Meissner effect}

\begin{figure}[ptb]
\begin{center}
\includegraphics[
height=5cm,
]
{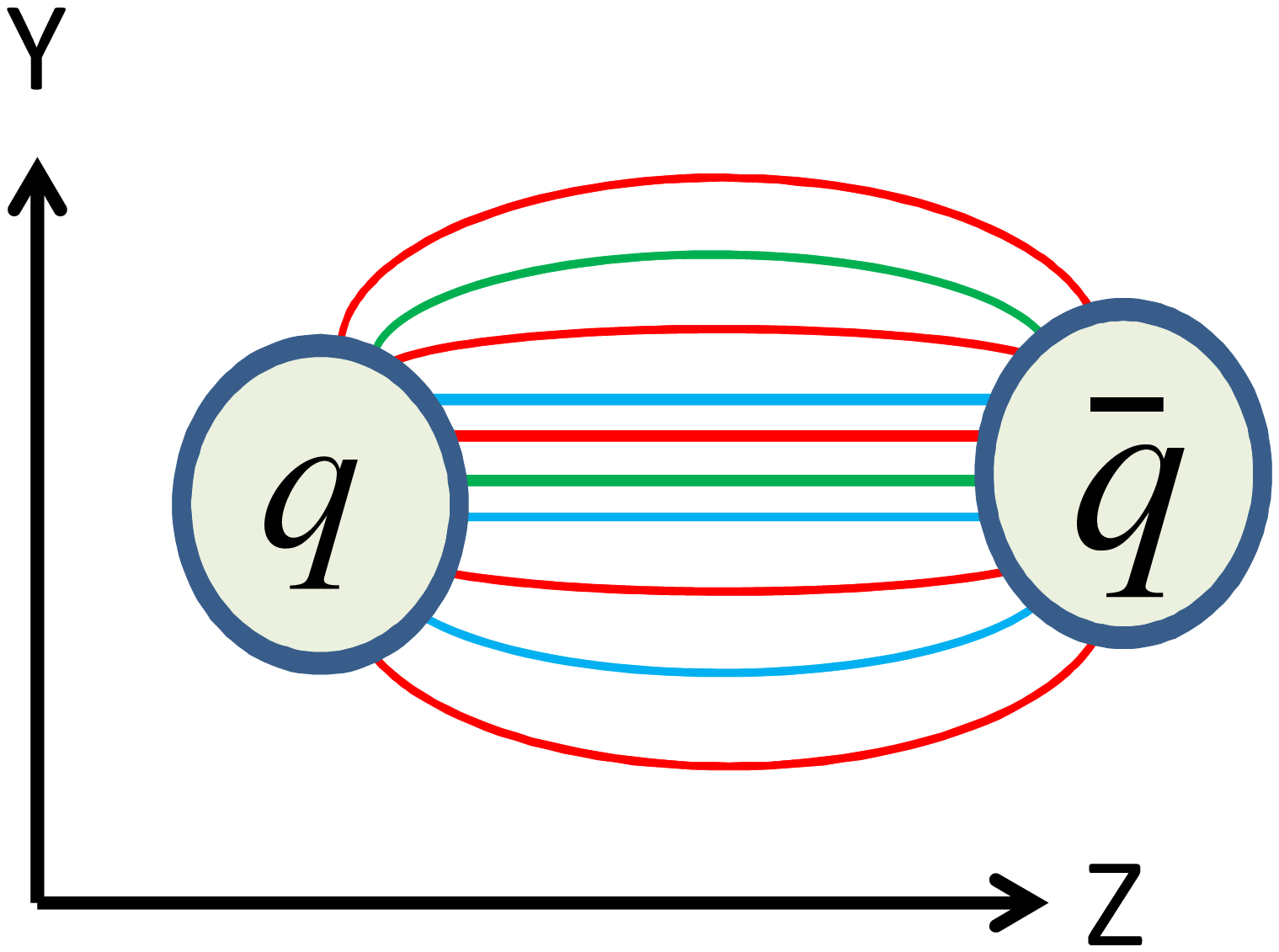} 
\ \ \ \ \ \ \ \ \ \ 
\includegraphics[
origin=c,
height=5.5cm,
]
{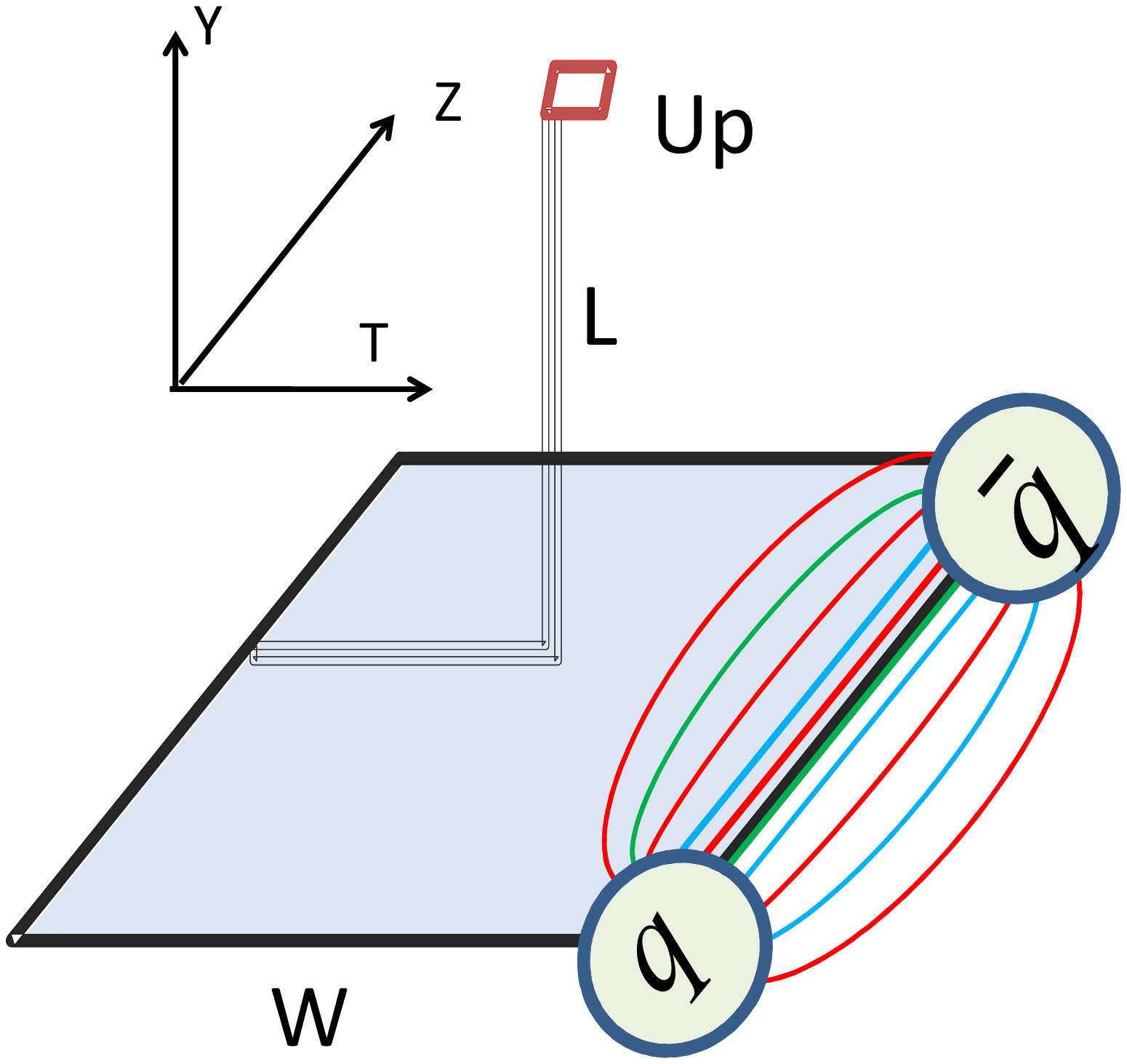}
\end{center}
\caption{(Left) 
The setup of measuring the chromo-flux produced by a quark--antiquark pair. 
(Right) 
The gauge-invariant connected correlator ($U_{p}LWL^{\dag})$ between a plaquette $U$ and the Wilson loop $W$.
}%
\label{fig:Operator}%
\end{figure}

We investigate the non-Abelian dual Meissner effect as the mechanism of quark confinement.
In order to extract the chromo-field, we use a gauge-invariant correlation function proposed in  \cite{DiGiacomo:1990hc}: 
The chromo-field created by a quark-antiquark pair in $SU(N)$ Yang-Mills theory is measured by using a gauge-invariant connected correlator between a plaquette and the Wilson loop  
(see Fig.\ref{fig:Operator}):%
\begin{equation}
\rho_{W}:=\frac{\left\langle \mathrm{tr}\left(  U_{P}L^{\dag}WL\right)
\right\rangle }{\left\langle \mathrm{tr}\left(  W\right)  \right\rangle
}-\frac{1}{N}\frac{\left\langle \mathrm{tr}\left(  U_{P}\right)
\mathrm{tr}\left(  W\right)  \right\rangle }{\left\langle \mathrm{tr}\left(
W\right)  \right\rangle }, \label{eq:Op}%
\end{equation}
where $W$ is the Wilson loop in
$Z$-$T$ plane representing a pair of quark and antiquark, $U_{P}$ a plaquette variable as the probe operator to measure the chromo-field strength at the point $P$, and $L$ the Wilson line connecting the source $W$ and the probe $U_{P}$.
Here $L$ is necessary to guarantee the gauge invariance of the correlator $\rho_{W}$ and hence the probe is identified with $LU_PL^\dagger$. 
The symbol $\left\langle \mathcal{O}\right\rangle $ denotes the average of the operator $\mathcal{O}$ in the space and the ensemble of the configurations. 
In the naive continuum limit $\epsilon \to 0$, indeed,  $\rho_{W}$ reduces to the field strength in the presence of the $q\bar q$ source:
\begin{equation}
\ \rho_{W}%
\overset{\varepsilon\rightarrow0}{\simeq}g\epsilon^{2}\left\langle
\mathcal{F}_{\mu\nu}\right\rangle _{q\bar{q}}:=\frac{\left\langle
\mathrm{tr}\left(  ig\epsilon^{2} L\mathcal{F}_{\mu\nu}L^{\dag}W \right)
\right\rangle }{\left\langle \mathrm{tr}\left(  W\right)  \right\rangle
}+O(\epsilon^{4}) ,
\end{equation}
where we have used $U_{x,\mu}=\exp (-ig\epsilon \mathcal{A}_\mu(x))$ and hence $U_{P}= \exp(-ig\epsilon^2 \mathcal{F}_{\mu\nu} )$.
Thus, the gauge-invariant chromo-field strength $F_{\mu\nu}[U]$ is given by 
\begin{equation}
F_{\mu\nu}[U] := \epsilon^{-2} \sqrt{\frac{\beta}{2N}}\rho_{W} ,
\end{equation}
where $\beta:=2N/g^2$ is the lattice gauge coupling constant. 
Note that the connected correlator $\rho_{W}$ is sensitive to the field strength, while  the disconnected one probes the squared field strength:
\begin{equation}
\rho_{W}^{\prime}:=\frac{\left\langle \mathrm{tr}\left(  W\right)
\mathrm{tr}\left(  U_{p}\right)  \right\rangle }{\left\langle \mathrm{tr}%
\left(  W\right)  \right\rangle }-\left\langle \mathrm{tr}\left(
U_{p}\right)  \right\rangle \overset{\varepsilon\rightarrow0}{\simeq}%
g\epsilon^{4}\left[  \left\langle \mathcal{F}_{\mu\nu}^{2}\right\rangle
_{q\bar{q}}-\left\langle \mathcal{F}_{\mu\nu}^{2}\right\rangle _{0}\right]  .
\end{equation}

\begin{figure}[ptb]
\begin{center}
\includegraphics[
width=6.2cm,
angle=270,
]
{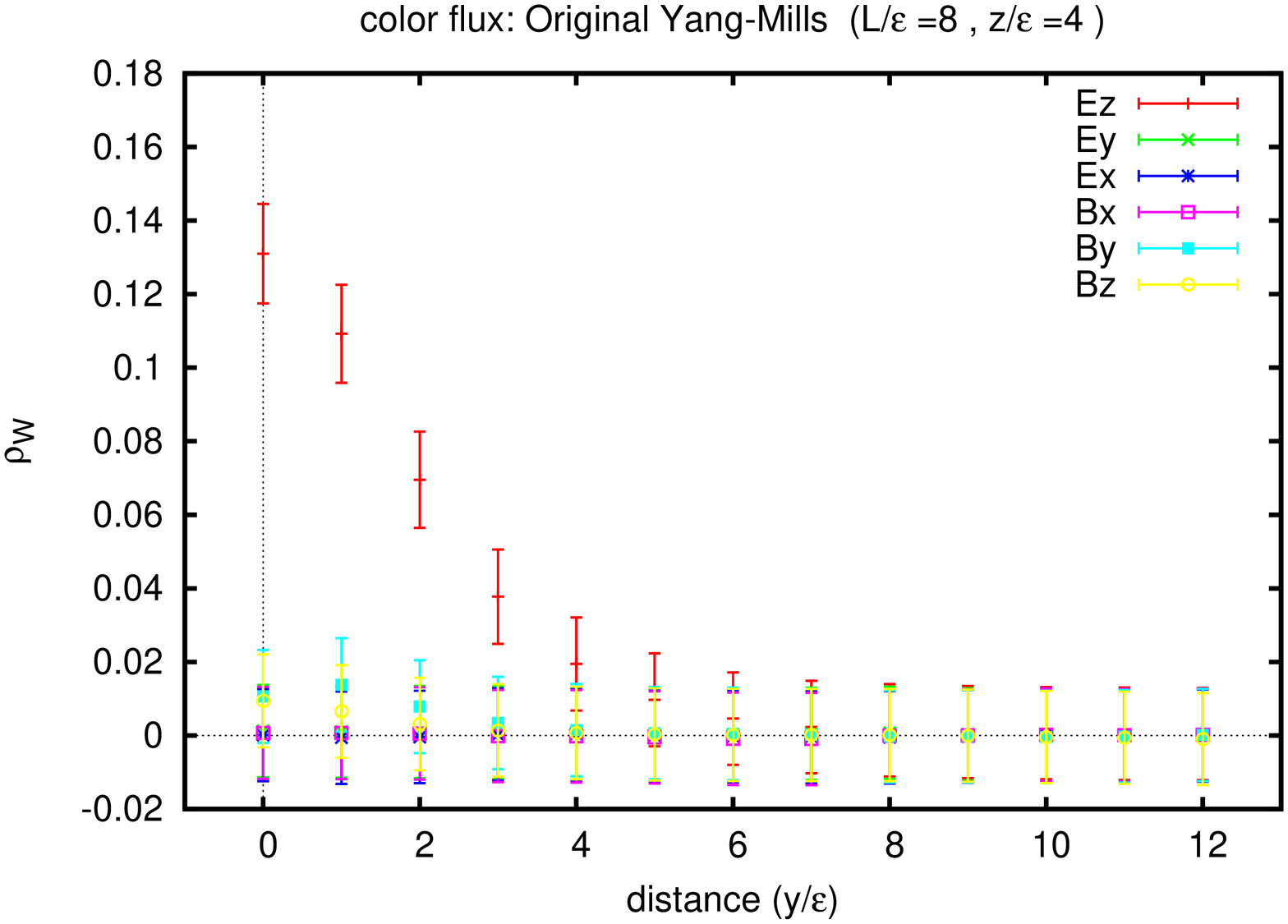}  
\ 
\includegraphics[
width=6.2cm,
angle=270,
]
{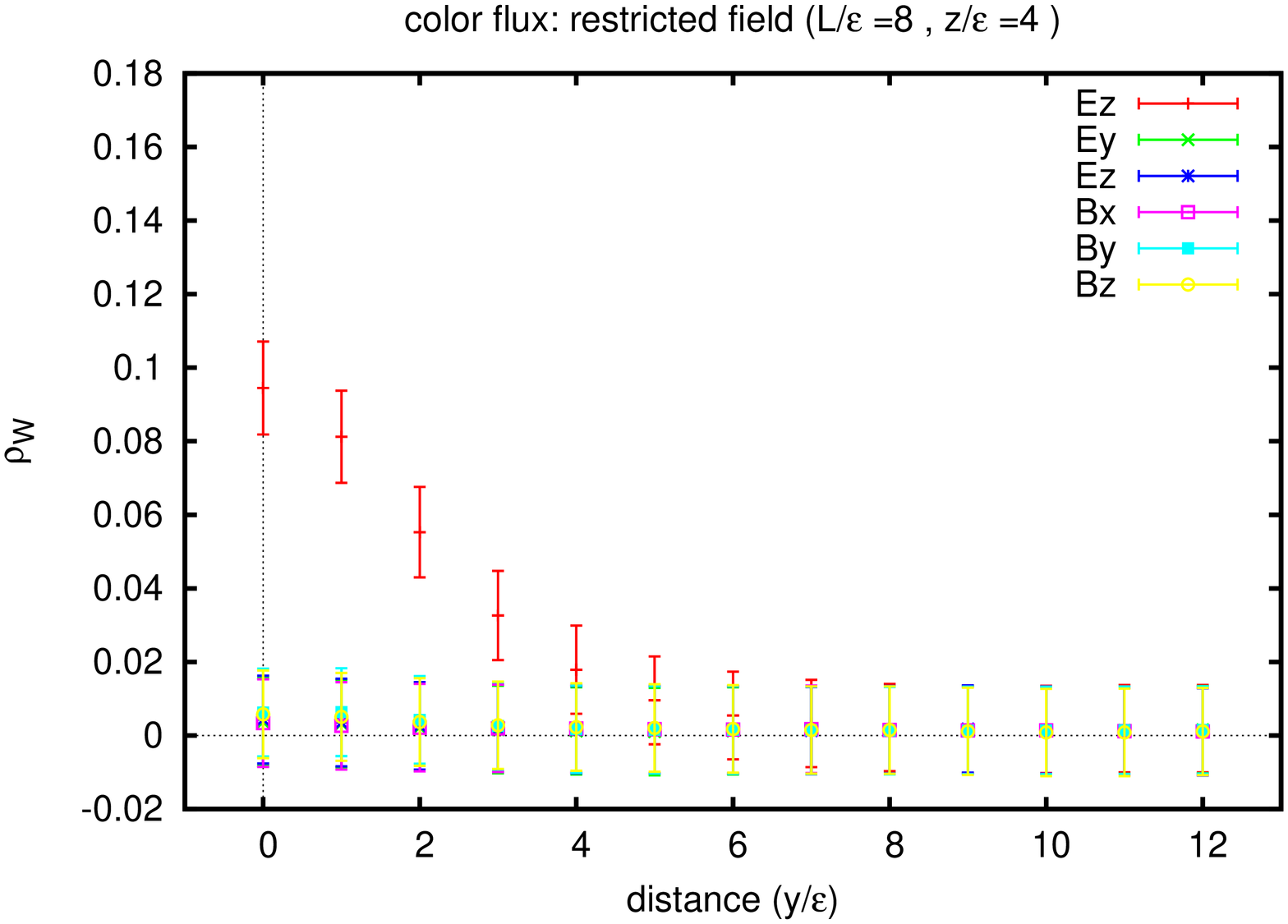}
\end{center}
\caption{
Measurement of components of the chromoelectric field $\bm{E}$ and chromomagnetic field  $\bm{B}$ as functions of the distance $y$ from the $z$ axis. 
(Left panel) the original $SU(3)$ YM field, 
(Right panel) the restricted $U(2)$ field. 
}%
\label{fig:measure}%
\end{figure}

We measure correlators between the plaquette $U_P$ and the chromo-field strength of the restricted field $V_{x,\mu}$ as well as the original YM field $U_{x,\mu}$.
See the left panel of Fig.~\ref{fig:Operator}.
Here the quark and antiquark source is introduced as $8\times8$ Wilson loop ($W$) in the $Z$-$T$ plane, and the probe $(U_{p})$ is set at the center of the Wilson loop and moved along the $Y$-direction. 
The left and right panel of Fig.~\ref{fig:measure} show respectively the results of measurements for the chromoelectric and chromomagnetic fields $F_{\mu\nu}[U]$ for the original  $SU(3)$ field $U$ and $F_{\mu\nu}[V]$ for the restricted  field $V$, where the field strength $F_{\mu\nu}[V]$  is obtained by using $V_{\,x,\mu}$ in eq(\ref{eq:Op}) instead of $U_{x,\mu}$:
\begin{equation}
F_{\mu\nu}[V] := \sqrt{\frac{\beta}{2N}} \tilde\rho_{W} , \quad 
\tilde\rho_{W}:=\frac{\left\langle \mathrm{tr}\left(  LV_{p}L^{\dag}W \right)
\right\rangle }{\left\langle \mathrm{tr}\left(  W\right)  \right\rangle
}-\frac{1}{N}\frac{\left\langle \mathrm{tr}\left(  V_{p}\right)
\mathrm{tr}\left(  W\right)  \right\rangle }{\left\langle \mathrm{tr}\left(
W\right)  \right\rangle } .  
\end{equation}
We have checked that even if  $W[U]$ is replaced by $W[V]$, together with replacement of the probe $LU_{P}L^\dagger$ by the corresponding $V$ version, the  change in the magnitude of the field strength $F_{\mu\nu}$ remains within at most a few \%.

From Fig.\ref{fig:measure} we find that  only the $E_{z}$ component of the chromoelectric field $(E_x,E_y,E_z)=(F_{10},F_{20},F_{30})$ connecting $q$ and $\bar q$ has non-zero value for both the restricted field $V$ and the original YM field $U$.
The other components are zero consistently within the numerical errors. 
This means that the chromomagnetic field $(B_x,B_y,B_z)=(F_{23},F_{31},F_{12})$ connecting $q$ and $\bar q$ does not exist  and that the chromoelectric field is parallel to the $z$ axis on which  quark and antiquark are located.
The magnitude $E_{z}$ quickly decreases in the distance $y$ away from the Wilson loop, as discussed in more detail in subsection C.

\begin{figure}[ptb]
\begin{center}
\includegraphics[
height=8.0cm,
angle=270
]
{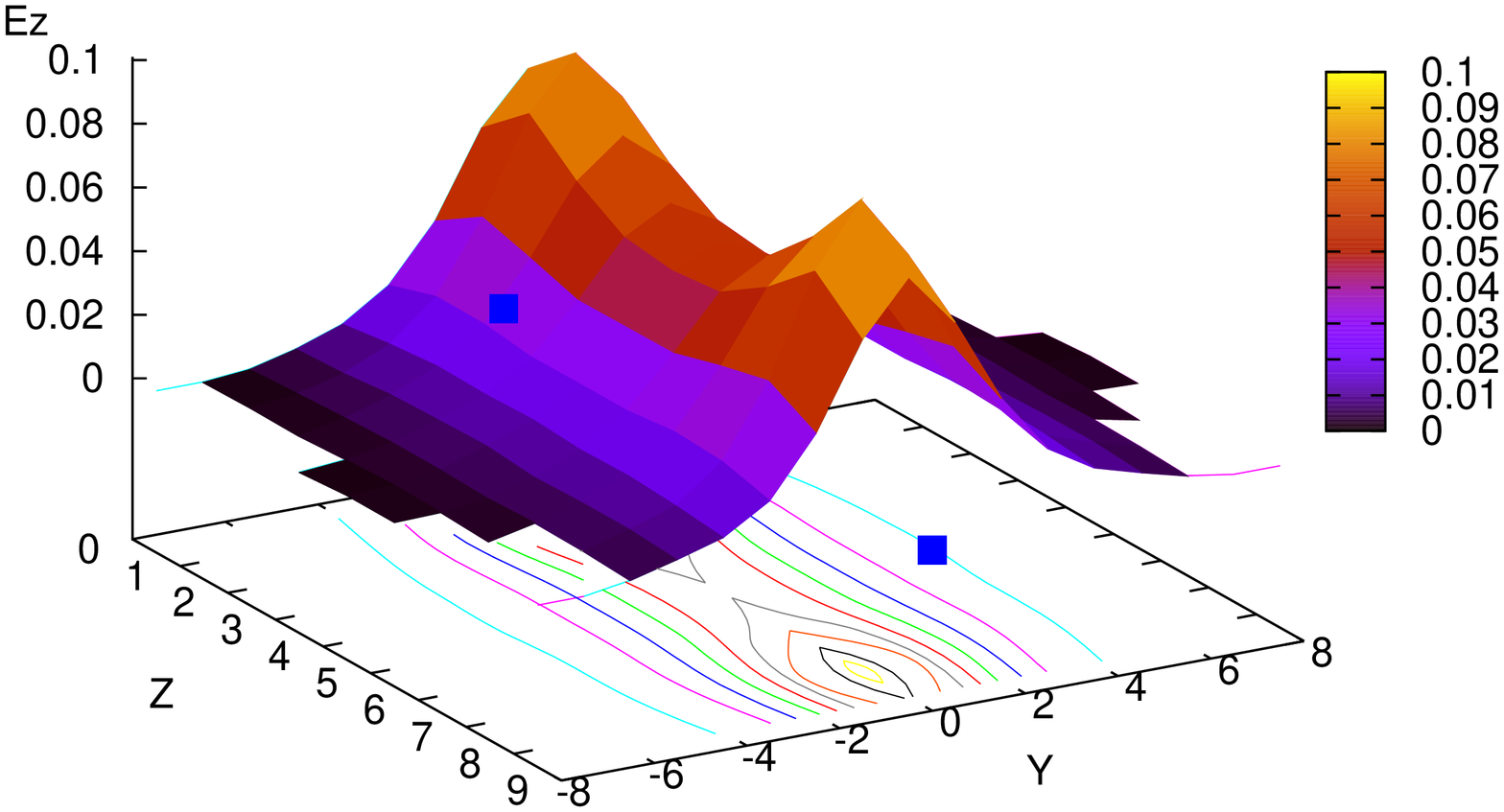} 
\quad
\includegraphics[
height=8.0cm,
angle=270
]
{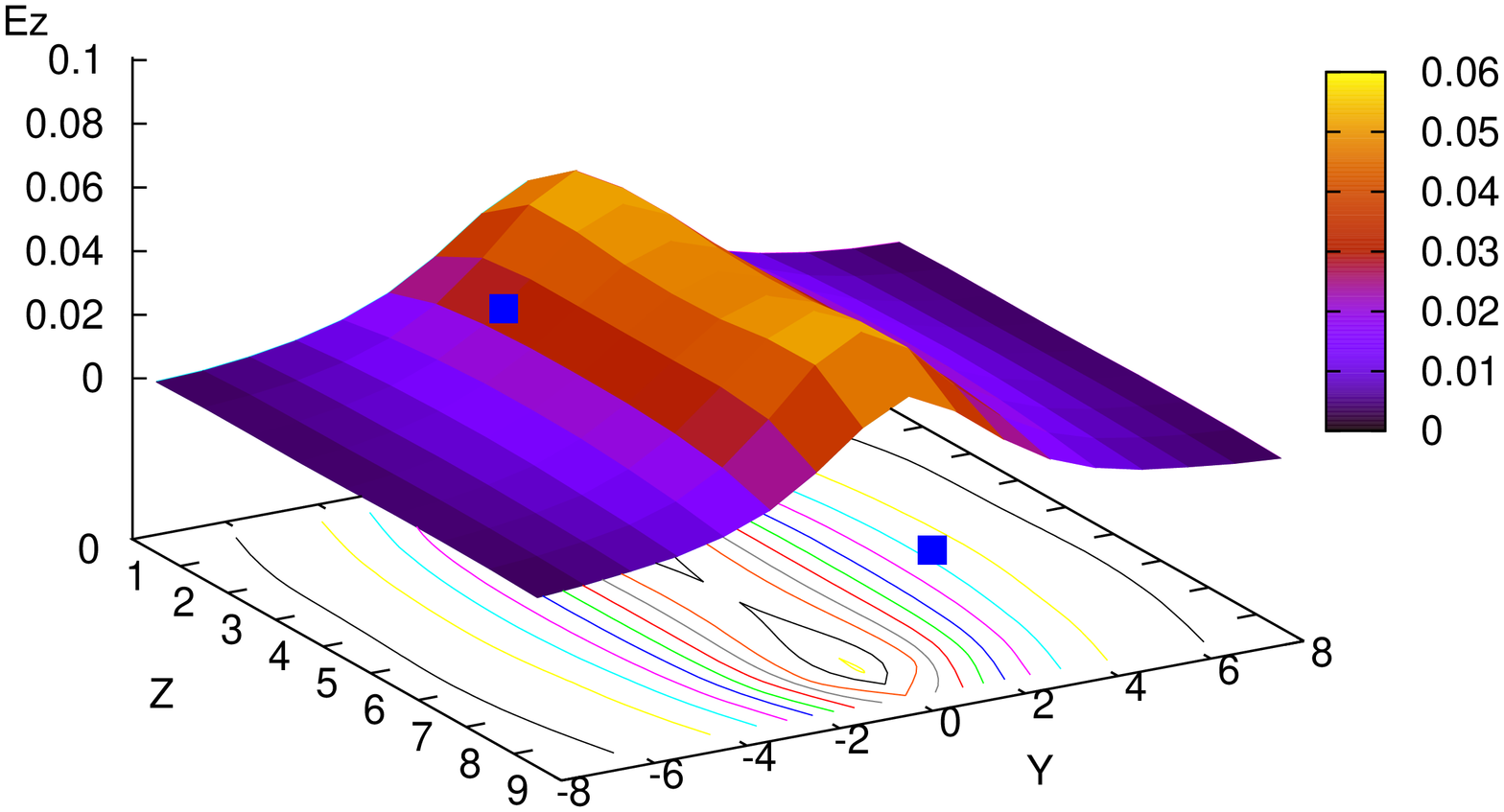} 
\vspace{-5mm}
\end{center}
\caption{
The distribution in $Y$-$Z$ plane of the chromoelectric field $E_z$ connecting a pair of quark and antiquark: 
(Left panel) chromoelectric field produced from the original YM field, 
(Right panel) chromoelectric field produced from the restricted $U(2)$ field. 
}%
\label{fig:fluxtube}%
\end{figure}

To see the profile of the nonvanishing component $E_z$ of the chromoelectric field in detail, we explore the distribution of chromoelectric field on the 2-dimensional plane. 
 Fig.~\ref{fig:fluxtube} shows the distribution of $E_{z}$ component of the chromoelectric field, where the quark-antiquark source represented as $9\times11$ Wilson loop $W$ is placed at $(Y,Z)=(0,0), (0,9)$, and the probe $U$ is displaced on the $Y$-$Z$ plane at the midpoint of the $T$-direction. 
 The position of a quark and an antiquark is marked by the solid (blue) box. The magnitude of $E_{z}$ is shown by the height of the 3D plot and also the contour plot in the bottom plane.
 The left panel of Fig.~\ref{fig:fluxtube} shows the plot of $E_z$ for the $SU(3)$ YM field $U$, and the right panel of Fig.~\ref{fig:fluxtube} for the restricted-field $V$. 
We find that the magnitude $E_{z}$ is quite uniform for the restricted part $V$, while it is almost uniform for the original part $U$ except for the neighborhoods of the locations of $q$, $\bar q$ source. 
This difference is due to  the contributions from the remaining part $X$ which affects only the  short distance, as will be discussed in the next section. 

\subsection{Magnetic current}

Next, we investigate the relation between the chromoelectric flux and the  magnetic current. 
The magnetic(-monopole) current can be calculated as
\begin{equation}
\mathbf{k}={}^{\ast}dF[\mathbf{V}] ,
\label{def-k}
\end{equation}
where $F[\mathbf{V}]$ is the field strength (2-form) of the restricted field (1-form) $\mathbf{V}$, $d$
the exterior derivative and $^{\ast}$ denotes the Hodge dual operation. 
Note that non-zero magnetic current follows from violation of the Bianchi identity  (If the field strength was given by
the exterior derivative of $\mathbf{V}$ field (one-form), $F[\mathbf{V}]=d\mathbf{V}$, \ we would obtain $\mathbf{k=}^{\ast}d^{2}\mathbf{V}$ $=0$).

\begin{figure}[ptb]
\begin{center}
\includegraphics[
width=6.5cm
]
{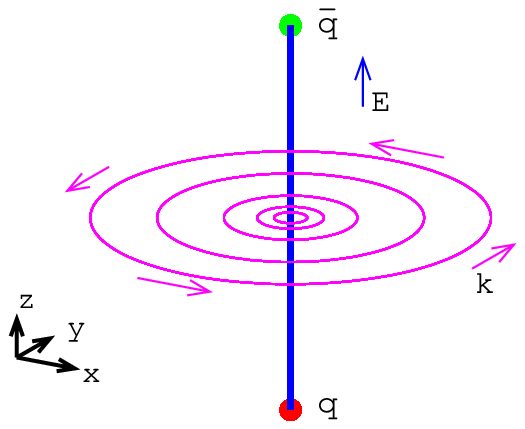} 
\quad
\includegraphics[
width=10.5cm
]
{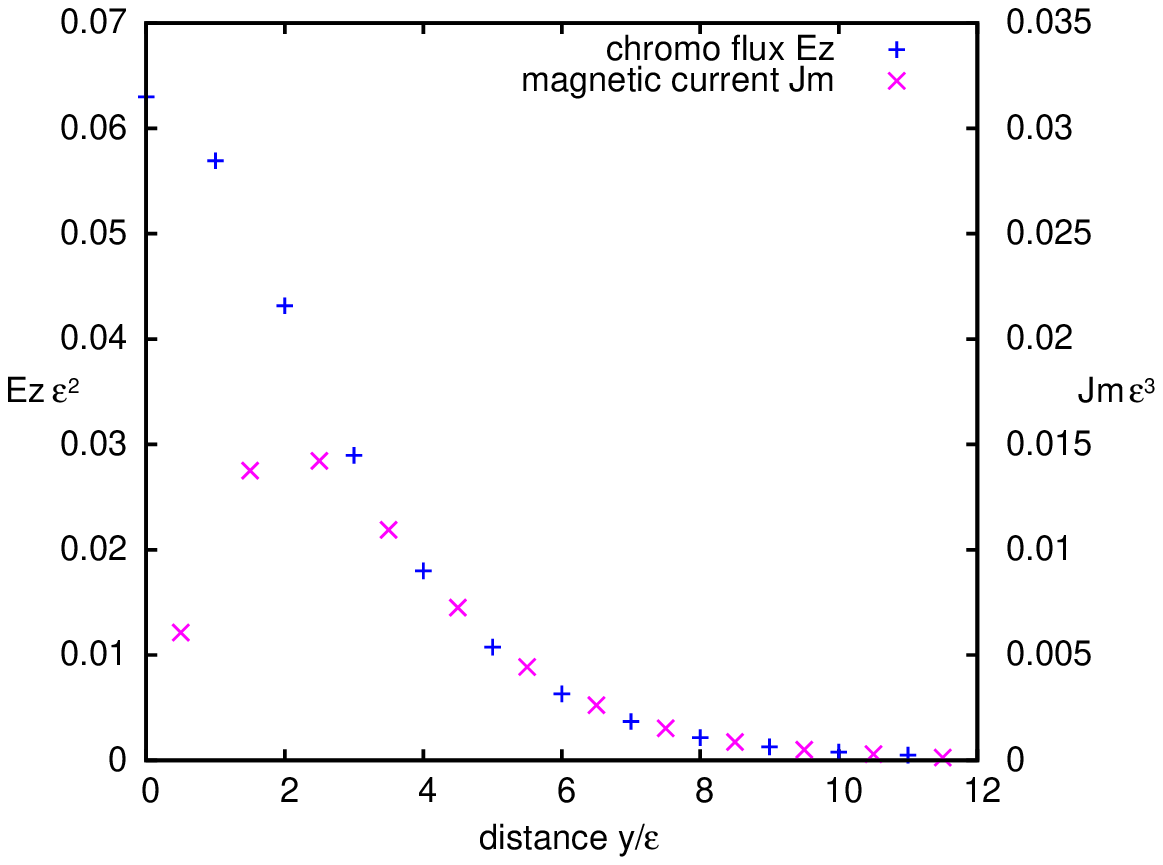} 
\vspace{-5mm}
\end{center}
\caption{{}The magnetic-monopole current $\mathbf{k}$ induced around the flux along the $z$ axis connecting a  quark-antiquark pair.
(Left panel) The positional relationship between the chromoelectric field $E_{z}$ and the magnetic current $\mathbf{k}$. 
(Right panel) The magnitude of the chromo-electronic current $E_{z}$ and the magnetic current  $J_{m}=|\mathbf{k}|$ as functions of the distance $y$ from the $z$ axis. 
}
\label{fig:Mcurrent}%
\end{figure}

Fig.~\ref{fig:Mcurrent} shows the  magnetic current measured in $X$-$Y$ plane at the midpoint of quark and antiquark pair in the $Z$-direction. 
The left panel of Fig.~\ref{fig:Mcurrent} shows the positional relationship between chromoelectric flux and  magnetic current.
The right panel of Fig.~\ref{fig:Mcurrent} shows the magnitude of the  chromoelectric field $E_z$ (left
scale) and the magnetic current $k$ (right scale). 
The existence of nonvanishing magnetic current $k$ around the chromoelectric field $E_z$ supports the dual picture of the ordinary superconductor exhibiting the electric current $J$ around the magnetic field $B$.

In our formulation, it is possible to define a gauge-invariant magnetic-monopole current  $k_{\mu}$  by using $V$-field,
which is obtained from the field strength $\mathcal{F}_{\mu\nu}[\mathbf{V}]$ of the field $\mathbf{V}$, as suggested from the non-Abelian Stokes theorem \cite{KondoNAST,KondoShibata}.
It should be also noticed that this
magnetic-monopole current  is a non-Abelian magnetic monopole extracted from the $V$ field, which corresponds to the stability group $\tilde{H}=U(2)$.
The magnetic-monopole current  $k_{\mu}$ defined in this way can be used to study the magnetic current around the chromoelectric flux tube, instead of the above definition of $k$  (\ref{def-k}).
The comparison of two monopole currents $k$ will be done in the forthcoming paper.

\subsection{Type of dual superconductivity}

\begin{figure}[ptb]
\begin{center}
\includegraphics[
width=6.0cm,
angle=270,
]
{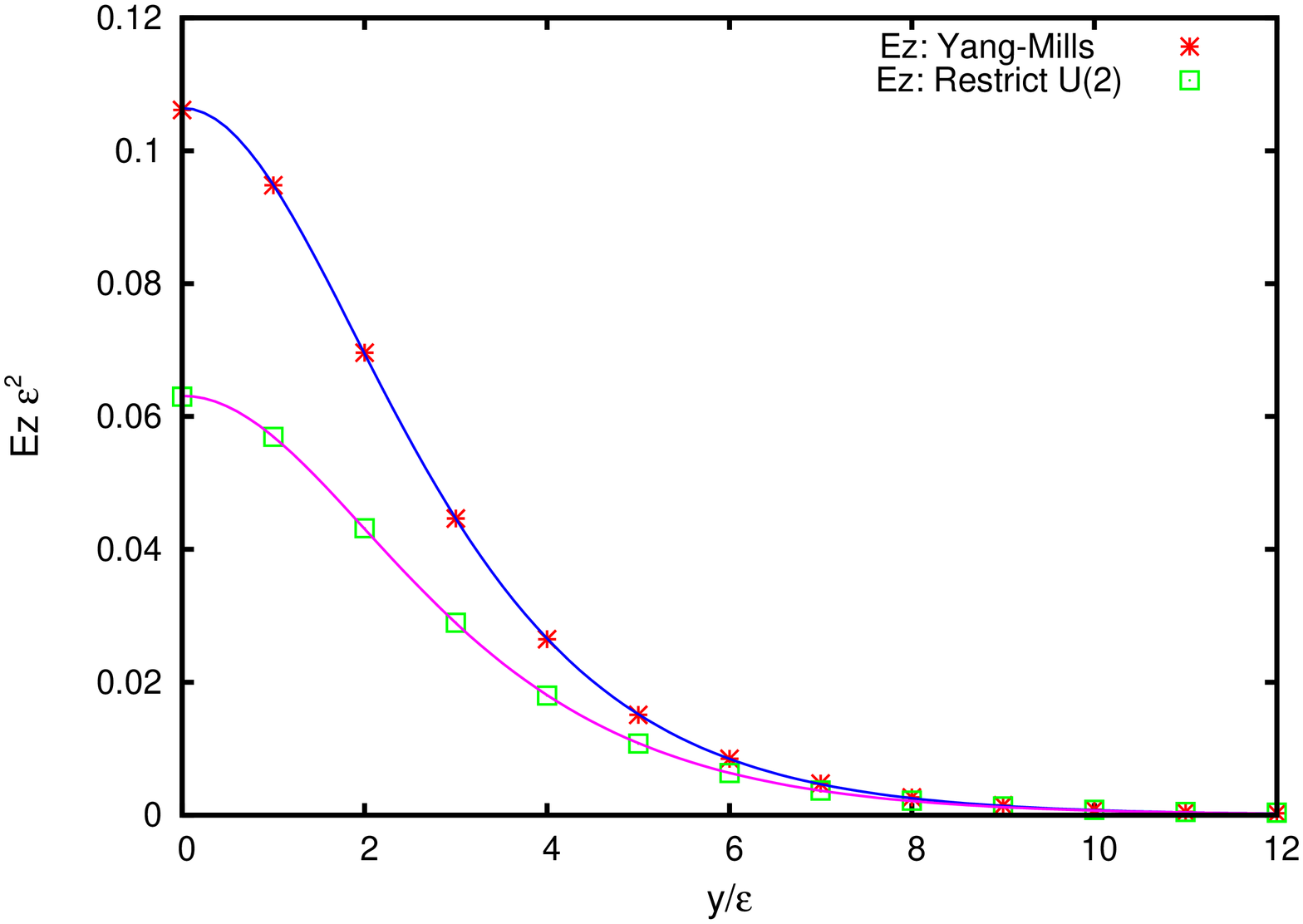} 
\includegraphics[
width=6.0cm,
angle=270,
]
{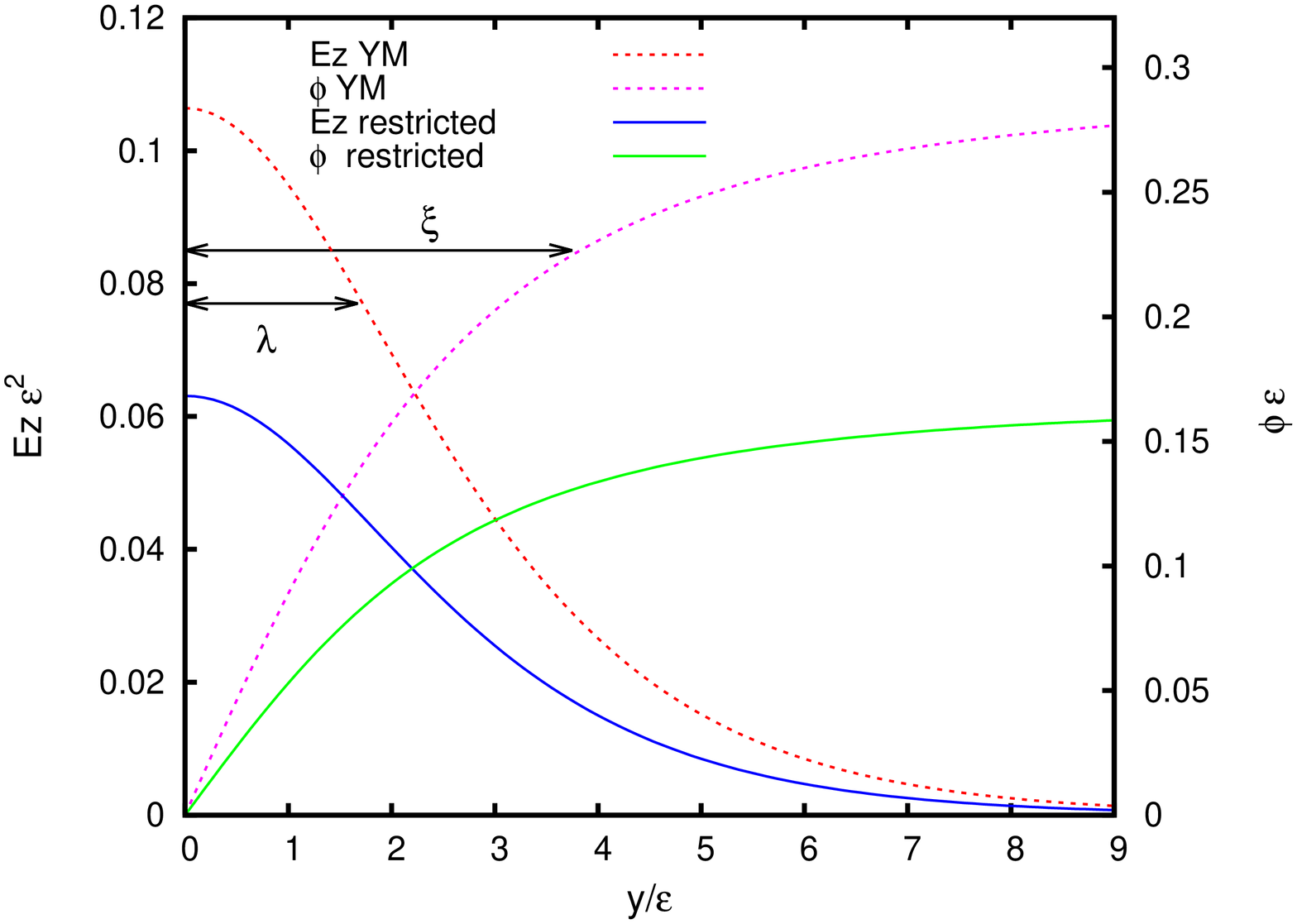}  
\vspace{-5mm}
\end{center}
\caption{
(Left panel)
The plot of the chromoelectric field $E_z$ versus the distance $y$ in units of the lattice spacing $\epsilon$ and the fitting   as a function $E_z(y)$ of  $y$ according to (\ref{fitting}). 
The red cross for the original $SU(3)$ field and the green square symbol for the restricted field.  
(Right panel) The order parameter $\phi$ reproduced as a function $\phi(y)$ of  $y$ according to (\ref{order-f}), togather with the chromoelectric field $E_z(y)$.
}
\label{fig:type}%
\end{figure}

Moreover, we investigate the QCD vacuum, i.e., type of the dual superconductor. 
The left panel of Fig.\ref{fig:type} is the plot for the chromoelectric field $E_z$ as a function of the distance $y$ in units of the lattice spacing $\epsilon$ for the original $SU(3)$ field and for the restricted  field.

In order to examine the type of the dual superconductivity, we apply the formula for the magnetic field  derived by Clem \cite{Clem75} in the ordinary superconductor based on the Ginzburg-Landau (GL) theory to the chromoelectric field in the dual superconductor.
In the GL theory, the gauge field $A$ and the scalar field $\phi$ obey simultaneously  the GL equation:
\begin{equation}
 (\partial^\mu -iq A^\mu)(\partial_\mu -iq A_\mu) \phi + \lambda (\phi^* \phi - \eta^2) = 0 ,
\end{equation}
and the Ampere equation:
\begin{equation}
 \partial^\nu F_{\mu\nu} + iq [\phi^* (\partial_\mu \phi -iq A_\mu \phi)  - (\partial_\mu \phi -iq A_\mu \phi)^* \phi] = 0 .
\end{equation}

Usually, in the dual superconductor of the type II, it is justified to use the asymptotic form $K_0(y/\lambda)$ to fit the chromoelectric field in the large $y$ region (as the solution of the Ampere equation in the dual GL theory).  
However, it is clear that this solution cannot be applied to the small $y$ region, as is easily seen from the fact that $K_0(y/\lambda) \to \infty$ as $y \to 0$. 
In order to see the difference between type I and type II, it is crucial to see the relatively small $y$ region.
Therefore, such a simple form cannot be used to detect the type I dual  superconductor. 
However, this important aspect was ignored in the preceding studies except for a work \cite{Cea:2012qw}.

On the other hand, Clem \cite{Clem75} does not obtain the analytical solution of the GL equation explicitly and use an approximated form for the scalar field  $\phi$ (given below in (\ref{order-f})).
This form is used to solve the Ampere equation exactly to obtain the analytical form for the gauge field $A_\mu$ and the resulting magnetic field $B$.
This method does not change the behavior of the gauge field in the long distance, but it gives a finite value for the gauge field even at the origin. 
Therefore, we can obtain the formula which is valid for any distance (core radius) $y$ from the  axis connecting $q$ and $\bar{q}$: the profile of chromoelectric field in the dual superconductor is obtained:
\begin{equation}
E_{z}(y)=\frac{\Phi}{2\pi}\frac{1}{\zeta\lambda}\frac{K_{0}(R/\lambda)}%
{K_{1}(\zeta/\lambda)},\text{ }R=\sqrt{y^{2}+\zeta^{2}} ,
\label{eq:fluxClem}%
\end{equation}
provided that the scalar field is given by (See the right panel of Fig.\ref{fig:type})
\begin{equation}
 \phi(y) = \frac{\Phi}{2\pi} \frac{1}{\sqrt{2}\lambda} \frac{y}{\sqrt{y^2+\zeta^2}} ,
 \label{order-f}
\end{equation}
where $K_{\nu}$ is the modified Bessel function of the $\nu$-th order,
$\lambda$ the parameter corresponding to the London penetration length, $\zeta$
a variational parameter for the core radius, and $\Phi$ external electric flux. 
In the dual superconductor, we define the GL parameter $\kappa$ as the ratio of the London penetration length $\lambda$ and the coherence length $\xi$ which measures the coherence of the magnetic monopole condensate (the dual version of the Cooper pair condensate):
\begin{equation}
\kappa= \frac{\lambda}{\xi}  .
\end{equation}  
It is given by \cite{Clem75}
\begin{equation}
\kappa=\sqrt{2} \frac{\lambda}{\zeta} \sqrt{1-K_{0}^{2}(\zeta/\lambda)/K_{1}^{2}(\zeta/\lambda)} .
\end{equation}

\begin{table}[t] 
\begin{tabular}{|l||c|c|c|| c|c|c|c|c|}\hline
& $a\epsilon^2$    & $b\epsilon$    & $c$       & $\lambda/\epsilon$ 
&$ \zeta/\epsilon$ & $\xi/\epsilon$ &  $\Phi$ & $\kappa$ \\
\hline
SU(3) YM field 
& $0.804 \pm 0.04$ & $0.598\pm 0.005$ & $1.878\pm 0.04$ & $1.672 \pm 0.014$ 
& $3.14\pm 0.09$ &  $3.75 \pm 0.12$  & $4.36 \pm 0.3$& $0.45 \pm 0.01$ \\
\hline
restricted field 
& $0.435 \pm 0.03 $ & $0.547 \pm 0.007$ & $1.787 \pm 0.05 $ & $1.828 \pm 0.023$ 
& $3.26 \pm 0.13$    & $3.84 \pm 0.19$    &$2.96 \pm 0.3 $     & $0.48 \pm 0.02$ \\
\hline
\end{tabular}
\caption{
The properties of the Yang-Mills vacuum as the dual superconductor obtained by fitting the data of chromoelectric field with the prediction of the dual Ginzburg-Landau theory.  
}
\label{Table:GL-fit}
\end{table}

According to the formula Eq.(\ref{eq:fluxClem}), we estimate the GL parameter $\kappa$ for the dual superconductor of $SU(3)$ YM theory, although this formula is obtained for the ordinary superconductor of $U(1)$ gauge field. 
By using the fitting function: 
\begin{equation}
E(y)=aK_{0}(\sqrt{b^{2}y^{2}+c^{2}}) , \quad
a= \frac{\Phi}{2\pi}\frac{1}{\zeta\lambda }\frac{1} {K_{1}(\zeta/\lambda)} ,
\quad
b=\frac{1}{\lambda}, 
\quad
c= \frac{\zeta}{\lambda}  ,
\label{fitting}
\end{equation}
we obtain the result shown in Table~\ref{Table:GL-fit}.
The superconductor is type I if $\kappa < \kappa_{c}$, while type II if $\kappa > \kappa_{c}$, where the critical value of GL parameter dividing the type of the superconductor is given by $\kappa_{c}=1/\sqrt{2}\simeq0.707$.
Our data clearly shows that the dual superconductor of $SU(3)$ Yang-Mills theory is type I with 
\begin{equation}
 \kappa=0.45 \pm 0.01 .
\end{equation} 
This result is consistent with a quite recent result obtained independently by  Cea, Cosmai and Papa \cite{Cea:2012qw}. 
Moreover, our result shows that the restricted  part plays the dominant role in determining the type of the non-Abelian dual superconductivity of the $SU(3)$ Yang-Mills theory, reproducing the same result \cite{ref:confinmentX}\cite{ref:SCGT12}, i.e., type I with  
\begin{equation}
 \kappa=0.48 \pm 0.02 .
\end{equation} 
This is a novel feature overlooked in the preceding studies. 
Thus the restricted-field dominance can be seen also in the determination of the type of dual superconductivity where the discrepancy is just the normalization of the chromoelectric field at the core $y=0$, coming from the difference of the total flux $\Phi$. 
These are  gauge-invariant results. 
Note again that this restricted-field and the non-Abelian magnetic monopole extracted from it reproduce the string tension in the static quark--antiquark potential \cite{lattice2010,abeliandomSU(3)}.

Our result should be compared with the result obtained by using the Abelian projection: Y.Matsubara et. al \cite{Matsubara:1993nq} suggests $\kappa=0.5 \sim 1$(which is $\beta$ dependent), border of type I and type II for both $SU(2)$ and $SU(3)$. 
In $SU(2)$ case, on the other hand, there are other works \cite{Suzuki:2009xy,Chernodub:2005gz} which conclude that the type of vacuum is at the border of type I and type II.


We should mention the work \cite{Cardoso:2010kw} which concludes that  the dual superconductivity of $SU(3)$ Yang-Mills theory is type II with $\kappa=1.2 \sim 1.3$.
This conclusion seems to contradict our result for $SU(3)$. 
If the above formula (\ref{eq:fluxClem}) is applied to the data of \cite{Cardoso:2010kw}, we have the same conclusion, namely, the type I with $\kappa=0.47 \sim 0.50$. 
Therefore, the data obtained in \cite{Cardoso:2010kw} are consistent with ours. 
The difference between type I and type II is attributed to the way of fitting the data with the formula for the chromo-field.

\section{Gluon propagators}

\begin{figure}[ptb]
\begin{center}
\includegraphics[
height=10.0cm
]
{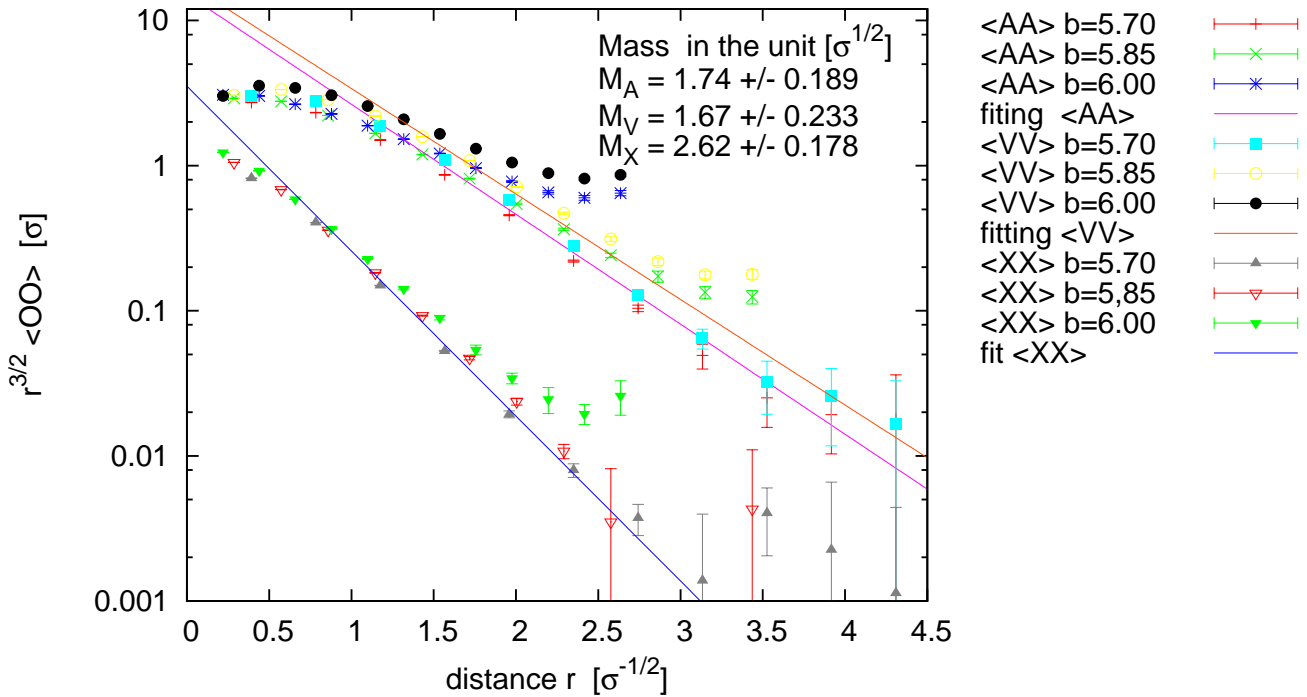} 
\vspace{-5mm}
\end{center}
\caption{
The rescaled correlation correlation functions $r^{3/2}\left\langle
O(r)O(0)\right\rangle$ for $O=\mathbf{A,V,X}$ for $24^{4}$ lattice with $\beta=5.7$, $5.85$, $6.0.$ 
The physical scale is set in units of the string tension $\sigma_{\text{phys}}^{1/2}$. 
The correlation functions have the profile of cosh type because of the periodic boundary condition, and hence we use  data within distance of the half size of lattice.
} 
\label{fig.propagator}%
\end{figure}

Finally, we study the 2-point correlation functions (propagators) of the new variables and the original Yang-Mills field variables, which are defined by
\begin{equation}
D_{OO}(x-y):=\left\langle O_{\mu}^{A}(x)O_{\mu}^{A}(y)\right\rangle \text{ for
}O_{\mu}^{A}(x )\in\{\mathbf{V}^{A}_{x^{\prime},\mu},\mathbf{X}^{A}
_{x^{\prime},\mu},\mathbf{A}^{A}_{x^{\prime},\mu}\},
\end{equation}
where an operator $O_{\mu}(x )$ is defined by the linear type, e.g.,
$\mathbf{A}_{x^{\prime},\mu}:=(U_{x,\mu}-U_{x,\mu}^{\dag})/(2ig\epsilon)$ where $x^{\prime}$ means the mid-point of $x$ and $x+\epsilon \mu$.
In order to calculate the propagators, we must impose a gauge fixing condition, and we have adopted the lattice Landau gauge (LLG).
Fig.~\ref{fig.propagator} shows the logarithmic plot of the scaled propagators $r^{3/2}D_{OO}(r)$ as a function of $r=|x-y|$, where the distance $r$ is drawn in units of the string tension $\sigma_{\text{phys}}$, and data of lattice spacing is taken from the TABLE I in Ref.\cite{Edward98}. 
The propagator $D_{VV}$ damps slowly and has almost the same damping behavior as $D_{AA}$, while the $D_{XX}$ damps quickly. Thus, from the viewpoint of the propagator, the $V$-field plays the dominant role in the deep infrared region or the long distance, while $X$-field is negligible in the long distance.
The rapid disappearance of $X$ contribution in the long distance is helpful to understand the difference of the profile of the flux tube in Fig.\ref{fig:fluxtube}. 

We estimate the mass $M_{O}$ (i.e., the rate of exponential fall-off) from the propagator $D_{OO}(r)$ by using the Fourier transformation of the massive
propagator in the Euclidean space, which  behaves for large $M_{O}r$ as
\begin{equation}
D_{OO}(r)=\int\frac{d^{4}k}{(2\pi)^{4}}e^{ik(x-y)}\frac{3}{k^{2}+M_{O}^{2}%
}\simeq\frac{3\sqrt{M_{O}}}{2(2\pi)^{3/2}}\frac{e^{-M_{O}r}}{r^{3/2}}\text{
\ \ (}M_{O}r\gg1\text{)},
\end{equation}
and hence the scaled propagator $r^{3/2}D_{OO}(r)$ should be proportional to $\exp(-M_{O}r)$. 
In order to perform the parameter fitting of $M_{O}$ for $O=\{\mathbf{V}_{x^{\prime},\mu},\mathbf{A}_{x^{\prime},\mu}\}$, 
we use data in the region
$[2.0,4.5]$ and exclude the data near the midpoint of the lattice to eliminate the finite volume effect, while for $O=\mathbf{X}_{x^{\prime},\mu}$ we use the region $[1.0,3.5].$ 
We use $\sigma_{\text{phys}}=440MeV$ to obtain preliminary result:
\begin{equation}
M_{A}\approx0.76  \text{ GeV, \ \ \ }M_{V}\simeq0.73\text{ GeV , \ \ \ }M_{X}%
\simeq1.15 \text{ GeV,}
\end{equation}
which should be compared with result of the maximal option \cite{lattce2007} in LLG, and also result of the Abelian projection in the maximal Abelian gauge \cite{Suganuma}.

\section{Summary and outlook}

In this paper, we have given further numerical evidences for confirming the non-Abelian dual superconductivity for $SU(3)$ YM theory proposed in \cite{abeliandomSU(3)}.
For this purpose, we have used our new formulation of $SU(3)$ YM theory on a lattice \cite{SCGTKKS08L,exactdecomp} to extract the restricted   field from the original $SU(3)$ YM field,  which has played a dominant role in confinement of quarks in the fundamental representation, i.e., the restricted-field dominance and the non-Abelian magnetic monopole dominance in the string tension, as shown in the previous studies \cite{abeliandomSU(3)}.

We have focused on the dual Meissner effect and have measured the chromoelectric field connecting a quark and an antiquark for both the original YM field and the restricted  field. 
We have observed the dual Meissner effect in $SU(3)$ YM theory, i.e., only the chromoelectric field exists and the magnetic-monopole current is induced around the flux connecting a quark and an antiquark. 
Moreover, we have determined the type of non-Abelian dual superconductivity, i.e., type I for $SU(3)$ YM theory, which should be compared with the border of type I and II for the SU(2) YM theory.
These features are reproduced only from the restricted  part. 
These results confirm the non-Abelian dual superconductivity picture for quark confinement.
 
 In order to confirm the type of the non-Abelian dual superconductivity, it will be interesting to investigate if  the   interactions between two chromoelectric flux tubes are attractive (expected for the  type I) or repulsive (expected for the type II).
 The interactions should reflect the internal non-Abelian nature which depends on the  distance between two tubes.
 It will be also interesting to extend our result to the finite temperature case to see the fate of the dual superconductivity,  and to see when and how the dual superconductivity disappears above the critical temperature.

\subsection*{Acknowledgement}

This work is supported by Grant-in-Aid for Scientific Research (C) 24540252
from the Japan Society for the Promotion Science (JSPS), and also in part by JSPS
Grant-in-Aid for Scientific Research (S) 22224003. The numerical calculations
are supported by the Large Scale Simulation Program No.T11-15 (FY2011) and
No.12-13 (FY2012) of High Energy Accelerator Research Organization (KEK).

\end{document}